# Record high superconducting transition temperature in $Ti_{1-x}Mn_x$ alloy with rich magnetic element Mn

Ying-Jie Zhang[#], Yijie Zhu[#], Qing Li*, Zhe-Ning Xiang, Tianheng Huang, Jian Sun* and Hai-Hu Wen*

National Laboratory of Solid State Microstructures and Department of Physics, Collaborative Innovation Center of Advanced Microstructures, Nanjing University, Nanjing 210093, China

[#]These authors contributed equally to this work.

*e-mail: liqing1118@nju.edu.cn; jiansun@nju.edu.cn; hhwen@nju.edu.cn

**It is well-known that magnetic moments are very harmful to superconductivity. A typical example is the element Mn whose compounds usually exhibit strong magnetism. Thus, it is very hard to achieve superconductivity in materials containing Mn. Here, we report enhanced superconductivity with the superconducting transition temperature ($T_c$) up to a record high-value of about 26 K in a beta-phase $Ti_{1-x}Mn_x$ alloy containing rich magnetic element Mn under high pressures. This is contrary to the intuition that the magnetic moments always suppress superconductivity. Under high pressures, we also found that in the middle-pressure regime, the Pauli limit of the upper critical field is surpassed. The synchrotron X-ray diffraction data shows an unchanged beta-phase with a continuous contraction of the cell volume, which is well supported by the first-principles calculations. Although the theoretical results based on electron-phonon coupling (EPC) can interpret the $T_c$ value in a certain pressure region, the monotonic enhancement of superconductivity by pressure cannot seek support from the theory. Our results show a surprising enhancement of superconductivity in $Ti_{1-x}Mn_x$ alloy with a considerable Mn content.**

The 3$d$ transition-metal based compounds always exhibit rich quantum states with novel phenomena, such as metal-insulator transitions, charge or spin density waves, quantum magnetism, superconductivity and so on [1-8]. In particular, the discoveries of unconventional high-temperature superconductivity (HTS) in cuprates and iron-based superconductors have stimulated enormous interests in compounds with 3$d$ electrons [9-14]. Even though unconventional superconductivity has been found in many compounds with 3$d$ transition-metal elements, in which the 3$d$ electrons dominate the occupied states near the Fermi Level [6, 7, 8, 15-17], it has been found to be relatively difficult to induce superconductivity in Chromium (Cr)- and Manganese (Mn)-based compounds, where the strong magnetism of these elements usually has an antagonistic effect on superconductivity [18-20]. Therefore, it is of great interest to explore possible superconductivity in Mn-based compounds. Actually, there have been some reports concerning the Mn-based superconductors, but the superconducting transition temperature $T_c$ was rather low, even under high pressures [21-26].

The discovery of pressure-induced superconductivity in MnP was reported in 2015 and it was recognized as the first Mn-based superconductor [21]. MnP undergoes a first transition from paramagnetic to ferromagnetic state at about 290 K, and then a second transition to double-helical order below 50 K [27]. By applying pressure, MnP shows superconductivity with a maximum value of 1 K in a narrow pressure range around 8 GPa [21]. Recently, a pressure-induced superconductivity above 9 K was reported in $A$Mn$_6$Bi$_5$ ($A$ = K, Rb, Cs) with a quasi-one-dimensional structure and the superconductivity was generated by suppressing the antiferromagnetic order [22, 25-26].

With the exceeding of the Pauli paramagnetic limit, authors suggested an unconventional magnetism-mediated pairing mechanism in $A$Mn$_6$Bi$_5$ series compounds. And it was the first ternary Mn-based superconductor with the highest $T_c$ than those of any other Mn-based superconductors. Besides, there are also some compounds with Mn that exhibit superconductivity under high pressures, such as MnSe and MnSb$_4$Te$_7$ [23, 24]. However, the highest $T_c$ reported so far in Mn-based superconductors is 9.5 K in RbMn$_6$Bi$_5$ achieved at about 15 GPa [25].

Although static magnetism is thought to be harmful to superconductivity [28-31], magnetic or spin fluctuations were regarded as an essential ingredient to mediate cooper pairing in unconventional superconductors [32, 33]. By applying chemical doping or high pressure, unconventional superconductivity with higher $T_c$ may be obtained in some compounds with rich magnetic element Mn. Keeping this important guideline in mind, we found that Matthias *et al.* reported a series of superconductive solid solutions with the addition of chromium, manganese, iron, or cobalt in titanium about 60 years ago [34]. They proposed that the significant enhancement of $T_c$ in Ti$_{1-x}$M$_x$ ($M$ = Cr, Mn, Fe, Co) alloy is much more than that of the usual alloying effects on the density of states or the variation of electron concentration. Our recent study on Ti$_{1-x}$Mn$_x$ clearly shows that the bulk superconductivity originates from the body-centered cubic ($\beta$) phase and the carrier density with a hole type is significantly reduced in the $\beta$ phase Ti$_{1-x}$Mn$_x$. More importantly, we observed possible evidence of magnetic or spin fluctuations from the magnetization measurements, which means that the magnetic moment of Mn may not induce an observable localized moment, but gives rise to enhanced superconductivity through possible magnetic interactions [35].

Pressure is a viable tuning parameter to modify the electronic and crystal structures in a clean and controllable way without introducing chemical impurities. In this work, we performed a high-pressure approach to seek higher $T_c$ in $Ti_{1-x}Mn_x$ by measuring its electrical transport properties under various pressures up to 160 GPa. Surprisingly, we found that the $T_c$ monotonically increases with increasing pressure and reaches a maximum of about 26 K under the pressure of 160 GPa in $Ti_{1-x}Mn_x$ with a considerable amount of Mn. The cell volume of the $\beta$ phase is continuously compressed without the occurrence of structural transition under high pressures up to 132 GPa. Moreover, we found that the upper critical field of $Ti_{1-x}Mn_x$ is significantly enhanced and surpasses the corresponding Pauli paramagnetic limit in the middle-pressure region. The calculated $T_c$ based on the VCA method agrees with the experimental value in a limited pressure region. However, in a wider pressure region, the quick enhancement of $T_c$ with increasing pressure cannot get an explanation from theoretical calculations based on the electron-phonon coupling picture. Combined with the violation of the Pauli paramagnetic limit in $Ti_{1-x}Mn_x$, we suggest an unconventional magnetism-mediated pairing mechanism in $Ti_{1-x}Mn_x$ under high pressures.

**$T_c$s and upper critical fields of $Ti_{1-x}Mn_x$ under high pressures.**

To investigate the superconducting behavior of $Ti_{1-x}Mn_x$ under different pressures, we performed electrical transport measurements in a diamond anvil cell (DAC) up to 160 GPa. Before the high-pressure resistance measurements, the sample quality of $Ti_{1-x}Mn_x$ at ambient pressure was characterized [35], which show bulk superconductivity with sharp transitions. Our high-pressure resistance measurements are performed on three $Ti_{0.88}Mn_{0.12}$ (S1-S3) and two $Ti_{0.84}Mn_{0.16}$ (S4, S5) samples, and all the samples crystallize into body-centered cubic structure. The experimental details about the sample classification, DAC apparatus, and pressure calibration can be found in the Methods, Supplementary Note 1 and Supplementary Fig. 1. Figure 1 (a) shows the temperature-dependent electrical resistance (*R-T*) for $Ti_{0.88}Mn_{0.12}$ (S2) from 2 K to 300 K under various pressures up to 80.3 GPa. It can be seen that the $T_c$ is enhanced under high pressures and the superconducting transition widths are still pretty narrow. Even though, the room temperature resistance and the normal state transport behavior of S2 are basically unchanged. To see the evolution of the superconducting transition more clearly, we normalized *R-T* curves with the resistance at 30 K and presented the data in Fig. 1(b). The value of $T_c$ increases from 2.5 K at ambient pressure to about 21.2 K at 80.2 GPa and shows no obvious saturation ($T_c$ here is referred to $T_c^{onset}$ defined as the temperature at which the resistance begins to deviate from the linear fitting of the normal state resistance). With further increasing the pressure up to 160 GPa, an onset superconducting transition up to about 26 K is achieved, as illustrated in Supplementary Figs. 2(a-b) and 3. This is the highest $T_c$ observed in compounds with rich magnetic element Mn and more than ten times compared to that of $Ti_{0.88}Mn_{0.12}$ at ambient pressure. The relatively broad superconducting transition width in Supplementary Figs. 2 and 3 can be attributed to the inhomogeneous pressure conditions within our high-pressure cell. This broadened transition behavior commonly occurs in high-pressure experiments using solid pressure-transmitting medium [36-38].

We also conducted electrical transport measurements on two $Ti_{0.84}Mn_{0.16}$ samples (S4, S5) to investigate the influence of more Mn concentrations on the $T_c$ in the title compound under high pressures. Figure 1(c) displays the *R-T* curves of a representative experiment run (S5) in a wide pressure range from 5.1 to 142 GPa and the corresponding enlarged view in the vicinity of the superconducting transition is shown in Fig. 1(d). With the increasing pressure, the $T_c$ shifts to a higher temperature with almost the same rate as $Ti_{0.88}Mn_{0.12}$ and finally reaches about 25 K at 142 GPa. We have performed a total of five runs of experiments to show the reproducibility of pressure-enhanced superconductivity in $Ti_{1-x}Mn_x$ alloy and the evolution of $T_c$ with pressure in different experiments is quite consistent with each other. Note that even though the superconductivity changes a lot under high pressures, the residual resistance ratio of $Ti_{1-x}Mn_x$ under high pressures is almost as small as it at ambient pressure, which may be due to the scattering from defects induced by the doping of magnetic elements [35, 39, 40].

To further characterize the superconducting state of $Ti_{1-x}Mn_x$ under high pressures, we investigate the response of superconductivity to magnetic field, and the experimental data are presented in Fig. 2(a-c) and Supplementary Figs. 4 and 5. We can see that the superconductivity is gradually suppressed and the $T_c$ of the pressurized sample shifts to low temperatures upon increasing magnetic fields. Besides, the normal-state resistance almost overlaps with each other, indicating that the magnetoresistance is negligible in $Ti_{1-x}Mn_x$ alloy. Surprisingly, we found that the superconductivity in $Ti_{1-}$

$_x$Mn$_x$ can maintain a relatively high transition temperature under a large magnetic field compared with element Ti and NbTi alloy [36, 38]. To quantify the evolution of the upper critical field ($\mu_0H_{c2}$), we have plotted the $\mu_0H_{c2}$-$T$ curves of Ti$_{1-x}$Mn$_x$ sample at three representative pressures in Fig. 2(d), here $T$ is referred to $T_c^{90\%}$ which is defined as the temperature at 90% of normal state resistance at different magnetic fields. The $\mu_0H_{c2}$-$T$ curve can be well fitted by the Ginzburg Landau (GL) equation

$$\mu_0 H_{c2}(T) = \mu_0 H_{c2}(0)[1-(T/T_c)^2]/[1+(T/T_c)^2] \qquad (1)$$

from which the upper critical field at 0 K ($\mu_0H_{c2}(0)$) could be estimated. The fitting results are shown as the colored solid lines in Fig. 2(d). The estimated values of $\mu_0H_{c2}(0)$ are 35.7 T (53.9 GPa), 48.2 T (80.2 GPa), and 42.4 T (142 GPa), and the corresponding Ginzburg-Landau coherence length $\xi_{GL}(0)$ is 3.03 nm, 2.61 nm, and 2.78 nm, respectively. To our surprise, we found that the $\mu_0H_{c2}(0)$ at 53.9 GPa and 80.2 GPa have exceeded the corresponding Pauli paramagnetic limit [41], implying the presence of strong coupling effect or a possible unconventional pairing mechanism by introducing magnetic element Mn into Ti$_{1-x}$Mn$_x$ alloy. However, under higher pressure (142 GPa), the estimated Pauli paramagnetic field will become larger than $\mu_0H_{c2}(0)$ as shown in Fig. 2(d). Moreover, we have measured additional $R$-$T$ curves of Ti$_{1-x}$Mn$_x$ under various magnetic fields at different runs and present the collected data in Supplementary Figs. 4 and 5. We can see that for both Ti$_{0.88}$Mn$_{0.12}$ and Ti$_{0.84}$Mn$_{0.16}$ samples, the magnetic fields can suppress superconductivity gradually and the fitting results of $\mu_0H_{c2}(0)$ show a similar pressure dependence. That is to say, accompanied with the increase of $T_c$, $\mu_0H_{c2}(0)$ of Ti$_{1-x}$Mn$_x$ increases rapidly and exceeds the Pauli limit firstly and finally falls behind.

**High-pressure synchrotron X-ray diffraction.**

To examine the stability of the crystal structure and whether the enhanced superconductivity in $Ti_{1-x}Mn_x$ alloy is accompanied with a structure phase transition, we performed high-pressure X-ray diffraction (HP-XRD) measurements on $Ti_{0.88}Mn_{0.12}$ sample at beamline BL15U1 of the Shanghai Synchrotron Radiation Facility, and the data is presented in Fig. 3(a). We can see that all the diffraction peaks continuously shift toward higher angles without new peaks appearing when the pressure increases up to 132 GPa, indicating the continuous compression of the lattice parameters and the absence of structural phase transition in $Ti_{1-x}Mn_x$ under high pressures. Figure 3(b) and (c) show the representative refinements at 132 GPa and 12.3 GPa. It is clear that all the diffraction peaks can be well indexed by a combination of the body-centered-cubic (bcc) structure (space group: $Im\bar{3}m$) of $Ti_{1-x}Mn_x$ and the elemental Re. Here, element Re exists as the calibration medium for HPXRD measurement. Combined with the equation of state (EOS) of Re under high pressures and the pressure dependence of the diamond Raman shift [42], we can calibrate the pressure value more accurately. We also show an additional set of high-pressure X-ray diffraction patterns of another $Ti_{0.88}Mn_{0.12}$ sample at a low-pressure region, the data is presented in Supplementary Fig. 6, and the results of two independent experiments show similar behavior under high pressures.

Figure 3(d) displays the pressure dependence of the calculated unit-cell volume (V). The cell volume of $Ti_{1-x}Mn_x$ shows a smooth and continuous decrease with increasing pressure and the overall volume decreases by about 40% at 132 GPa without volume collapse. Then we fit the pressure-dependent cell volume of $Ti_{0.88}Mn_{0.12}$ sample by the third-order Birch-Murnaghan (B-M) equation [43],

$$P(V) = \frac{3}{2} B_0 [(\frac{V_0}{V})^{7/3} - (\frac{V_0}{V})^{5/3}] \times \{1 + \frac{3}{4}(B_0' - 4)[(\frac{V_0}{V})^{2/3} - 1]\} \quad (2)$$

where $V_0$, $B_0$, and $B'$ are the unit cell volume at zero pressure, bulk modulus, and first-order derivative of the bulk modulus. The fitting results show $V_0 = 34.14$ Å$^3$, $B_0 = 104.03$ GPa, and $B' = 3.55$, respectively. The above results, including the absence of clear peak splitting and smooth contraction of unit-cell volume with pressure rule out the possible existence of structural transition in $Ti_{1-x}Mn_x$ under high pressures.

**Superconducting phase diagram of pressurized $Ti_{1-x}Mn_x$.**

Based on the above results, we construct a temperature-pressure (*T-P*) phase diagram as shown in Fig. 4(a). The value of $T_c$ is determined by the derivative of resistance *dR/dT* with respect to temperature. We find that $T_c$ firstly increases monotonously upon compression of the lattice, from 2.5 K at ambient pressure to about 24 K at around 100 GPa, then the increase rate $dT_c/dP$ reduces significantly. After reaching 130 GPa, $T_c$ increases at a relatively small rate, and finally reaches a maximum of 26 K at around 160 GPa, about ten times that observed at ambient pressure. We also extracted the pressure-dependent phase transition and $T_c$s of element Ti from previous literature [38, 44] and show the data in Fig. 4(b) and Fig. 4(a) (shown as the black dotted line) for comparison. The element Ti experiences a series of structural transitions under high pressures, $T_c$ undergoes a steep rise in the pressure range from 80 to 110 GPa and achieves a value of 26.2 K in the $\delta$ phase at an extreme high pressure of about 248 GPa [38]. Moreover, under higher pressures, the $T_c$ of $\beta$-Ti decreases continuously with increasing pressure due to the drastic phonon hardening [44]. However, the $Ti_{1-x}Mn_x$ alloy shows robust superconductivity with positive pressure dependence under high pressures throughout the $\beta$ phase, and the $T_c$ of $Ti_{1-x}Mn_x$ is much higher than that of Ti in the same pressure region, especially in the middle-pressure range from about 30 to 120 GPa.

In concomitant with the enhancement of $T_c$ in $Ti_{1-x}Mn_x$ under high pressures, the upper critical field at zero temperature $\mu_0H_{c2}(0)$ shows a nonlinear increment as illustrated in Fig. 4(c), that is, with the increase of pressure, it firstly increased rapidly, then gradually saturated, and finally showed a declining trend. We also plotted the $\mu_0H_{c2}(0)$ of element Ti at the same pressures for comparison, it is clear that the $\mu_0H_{c2}(0)$ of $Ti_{1-x}Mn_x$ is notably higher than those of element Ti, NbTi [36, 38, 44, 45], and high-entropy alloys [46, 47]. In addition, the pressure-dependent $\mu_0H_{c2}(0)/T_c$ along with the Pauli limit are given in Fig. 4(d). The $\mu_0H_{c2}(0)/T_c$ - $P$ relations of $Ti_{1-x}Mn_x$ show a broad dome behavior with a maximum value of about 2.6 at about 50 GPa. That is to say, the $\mu_0H_{c2}(0)/T_c$ of $Ti_{1-x}Mn_x$ is rapidly raised with increasing pressure and exceeds the Pauli limit firstly, and finally falls behind under higher pressures. The violation of the Pauli paramagnetic limit in a wide pressure range may indicate a possible strong electron correlation or unconventional superconductivity in $Ti_{1-x}Mn_x$ alloy. Moreover, we can see that a small variation of Mn content (from $x$ = 0.12 to 0.16) does not change the superconducting behavior of $Ti_{1-x}Mn_x$, including the pressure-dependent $T_c$ and

$\mu_0H_{c2}(0)$. A slight variation of the obtained value for different experimental runs may be due to the nonhydrostatic pressure conditions in our DAC chamber since we use solid pressure-transmitting medium.

**Theoretical calculations.**

To further understand the high $T_c$ in $Ti_{1-x}Mn_x$ under high pressures, we performed the first-principles calculations based on the VCA method. Firstly, we investigated the thermodynamic stability of the $\beta$ phase $Ti_{1-x}Mn_x$ and found that the calculated doping-dependent enthalpy differences between $\alpha$ phase and $\beta$ phase and pressure-dependent lattice parameters are in good agreement with the previous work [35] and experimental diffraction data as shown in Supplementary Fig. 7. Figure 5(a) presents the calculated phonon spectrum, phonon density of states, Eliashberg spectral function $\alpha^2F(\omega)$, and accumulated EPC strength $\lambda(\omega)$ of $Ti_{0.88}Mn_{0.12}$ at 100 GPa. The absence of imaginary phonon modes indicates that the compound is dynamically stable at this pressure. Additional phonon calculations establish the stability range to be between 60 and 110 GPa. The linewidth of the phonon spectrum is proportional to the relative contribution to the total $\lambda(\omega)$, and we found that the phonons below 100 cm$^{-1}$ contribute significantly to $\lambda$. In particular, the phonon branches along the N-Γ path make the largest contributions.

The calculated pressure-dependent total EPC constant $\lambda$, logarithmic frequency $\omega_{\log}$, and Debye temperature $\Theta_D$ of $Ti_{0.88}Mn_{0.12}$ in the pressure range from 60 to 110 GPa are summarized in Figs. 5(b) and (c). Details about the calculation of the above values can be seen in Supplementary Note 2. As the pressure increases, $\lambda$ shows a slow reduction below 80 GPa and then increases with pressure, whereas $\omega_{\log}$ exhibits the exact opposite pressure dependent. The calculated $\varTheta_D$ continuously increases with pressure. It is well known that the EPC strength and the density of state at the Fermi level ($N(E_F)$) are important for the enhancement of $T_c$ for a conventional phonon-mediated superconductor. We also calculated the electronic band structures of $Ti_{1-x}Mn_x$ under high pressures and present the results in Fig. 5(d) and Supplementary Figs. 8 and 9. The results show that the band structures hardly change with pressure and $N(E_F)$ slightly decreases with increasing pressure (see Fig. S8 and S9). By assuming different Coulomb repulsion constant $\mu^*$, we can theoretically predict the $T_c$ values of $Ti_{1-x}Mn_x$ in a short period region of pressure, as plotted in Fig. 5(e) and Supplementary Fig. S10. It is clear that the $T_c$ hardly changes under high pressures. For comparison, we also extracted experimental $T_c$ from transport measurements (yellow area) and put it in Fig. 5(e). To our surprise, although the theoretically predicted $T_c$ can agree with the experimental value in a narrow pressure range, the theoretical calculation cannot give consistent values in a wider pressure range, especially in the area where $T_c$ rises rapidly.

**Discussions**

From the above results, we know that the application of pressure can strongly enhance the superconducting transition temperature and the upper critical field of $Ti_{1-x}Mn_x$. To our knowledge, in $Ti_{1-x}Mn_x$, the superconducting properties, including the $T_c$s and $\mu_0H_{c2}$, are the highest among the transition-metal alloys, such as NbTi and other high-entropy alloy (HEA) superconductors [36, 47]. As we know, by applying pressure, a remarkable enhancement of $T_c$ can be observed in some transition metal elements, along with a series of structural transitions [38, 44, 45, 48-50]. And the evolution of $T_c$ in these elements can be explained by the enhanced electron-phonon coupling of conventional phonon-mediated superconductivity, which is closely related to pressure-induced *s-d* charge transfer [38, 44-45, 48-53]. Especially, in element Ti, accompanied by a series of structural transition sequences ($\alpha \to \omega \to \gamma \to \delta \to \beta$) under high pressures, dramatic pressure-enhanced superconductivity with the maximal $T_c$ around 23-26 K is achieved in $\delta$ phase [38, 44, 45], and the superconductivity is found to be significantly influenced by the *s-d* transfer and *d*-band driven correlation effects [38, 44]. By doping an appropriate amount of Mn into Ti, in this work, we obtained the $Ti_{1-x}Mn_x$ alloy with $\beta$ phase and found the monotonic increment of $T_c$ under high pressures. The structure of $Ti_{1-x}Mn_x$ still maintains the bcc structure without structural phase transition in the pressure range up to 132 GPa. This is completely contrary to the properties of Ti, in which the $\beta$ phase can only be stabilized above 200 GPa, and the $T_c$ decreases with increasing pressure [38, 44]. Combined with the unexpectedly large upper critical fields as shown in Fig. 4, we may expect an unconventional superconducting pairing mechanism in $Ti_{1-x}Mn_x$ under high pressures.

Usually, the compounds with 3*d* magnetic transition metal are commonly believed to be antagonistic to superconductivity and thus should have a low $T_c$, and in those compounds the upper critical fields of the superconducting states are intended to exceed the Pauli paramagnetic limit, indicating an unconventional pair mechanism [8, 22, 25-26,

[54]. In our experiments, the $T_c$ of $Ti_{1-x}Mn_x$ can be raised at least to tens of Kelvin under high pressures and the $\mu_0H_{c2}(0)$ can also violate the corresponding Pauli paramagnetic limit. The transition temperatures of $Ti_{1-x}Mn_x$ under high pressures are undoubtedly the highest among the superconductors that contain rich magnetic element Mn since the previous highest value is 9.5 K at 15 GPa in $RbMn_6Bi_5$ [25]. In our previous work, we have found possible evidence of antiferromagnetic spin fluctuations in $Ti_{1-x}Mn_x$ with $\beta$ phase and a strong electron correlation as evidenced by the relatively large Wilson ratio [35]. Usually, the strong correlation might cause an enhancement of the electronic effective mass ($m^*$) and a reduction of $g$ factor [55], which would result in a large upper critical field. Furthermore, the theoretical calculations based on the EPC mechanism also show that the computed $T_c$s cannot agree well with the experimental measurements over the full pressure range up to 160 GPa. Therefore, we may expect some unconventional origins for the enhanced $T_c$ and $\mu_0H_{c2}(0)$ in $Ti_{1-x}Mn_x$ under high pressures.

In conclusion, we have observed an unexpectedly high $T_c$ superconductivity at about 26 K in pressurized $Ti_{1-x}Mn_x$ alloy, this is a record high value of $T_c$ among compounds with rich magnetic element Mn. With the dramatic enhancement of $T_c$ under high pressures, the crystal structure of $Ti_{1-x}Mn_x$ consistently maintains the body-centered $\beta$ phase, and $\mu_0H_{c2}(0)$ is rapidly raised and exceeds the corresponding Pauli limit firstly and finally falls behind. The theoretical calculations show that the EPC scenario cannot fully interpret the evolution of $T_c$s of $Ti_{1-x}Mn_x$ under high pressures, indicating a possible unconventional pairing mechanism. Our finding not only provides fresh experimental data for a better understanding of the superconducting mechanism of transition metal alloys with high $T_c$ and $\mu_0 H_{c2}$, but also opens a new platform for finding more Mn-based superconductors with higher transition temperatures, and designing high-$T_c$ and high-$\mu_0 H_{c2}$ alloy superconductors for potential applications at extreme conditions.

**Methods**

**Sample preparation and High-pressure Resistance measurements.** High-quality single-phase $Ti_{1-x}Mn_x$ alloy was prepared by arc-melting method [35]. High pressure was generated by a DAC made of BeCu alloy with two opposing anvils. A four-probe van der Pauw method with platinum foil as electrodes was applied for resistance measurements. Diamond anvil cells (DACPPMS-ET225, Shanghai Anvilsource Material Technology Co., Ltd) with 300, 200, and 100 μm culets were used to generate high pressures up to 50, 80, and 160 GPa. Electrical transport measurements under high pressures were carried out on PPMS-9T (Quantum Design) in a temperature range from 2 to 300 K. Pressure below and above 80 GPa were determined by the ruby fluorescence method [56] and the pressure dependence of the diamond Raman shift method [42], respectively.

**High-pressure XRD measurements.** *In situ* high-pressure XRD measurements were carried out on the BL15U1 beamline at the Shanghai Synchrotron Radiation Facility. A monochromatic X-ray beam with a wavelength of 0.6199 Å was adopted for all measurements. Symmetric DACs with anvil culet sizes of 100 μm were used to generate pressure up to 132 GPa. Silicon oil was used as the pressure transmitting medium (PTM) and pressure was determined by the pressure dependence of the diamond Raman shift method [42]. The Rietveld refinements were conducted with TOPAS 4.2 software [57].

**Theoretical Calculations.** The VCA and superconducting properties of $Ti_{1-x}Mn_x$ were performed with density functional theory, as implemented in the Quantum ESPRESSO (QE) [58, 59]. The exchange-correlation potential is described with the norm-conserving pseudopotential based on the local density approximation (LDA) [60]. More details of our theoretical methods are described in Supplementary Note 2.

**Data availability**

Source data and all other data that support the plots within this paper and other finding of this study are available from the corresponding author upon reasonable request.

**Acknowledgments**

We thank Lili Zhang for her kind help during the HPXRD measurements at Shanghai Synchrotron Radiation Facility (SSRF). This work is financially supported by the National Key R&D Program of China (No. 2022YFA1403201), the National Natural Science Foundation of China (Nos. 12204231, 11927809, 12125404, and 12061131001), the Strategic Priority Research Program (B) of Chinese Academy of Sciences (No. XDB25000000), and the Fundamental Research Funds for the Central Universities. The high-pressure XRD measurements were performed on the beamline BL15U1 at Shanghai Synchrotron Radiation Facility (SSRF). The authors thank the support of the User Experiment Assist System of SSRF for *in situ* high-pressure Raman

**Author contributions**

H.-H. W., and Q. L. conceived and designed the experiments. Y.-J. Z., and Q. L. grew and characterized the samples. Q. L., Y.-J. Z., and Z.-N. X. performed high-pressure electrical transport and *in situ* XRD measurements with assistance from H.-H. W. Y. Z., T. H., and J. S. performed the theoretical calculations and analyses. H.-H. W., Q. L., and Y.-J. Z. analyzed the experimental data and wrote the manuscript with the inputs from all co-authors.

**Competing interests**

The authors declare no competing interests.

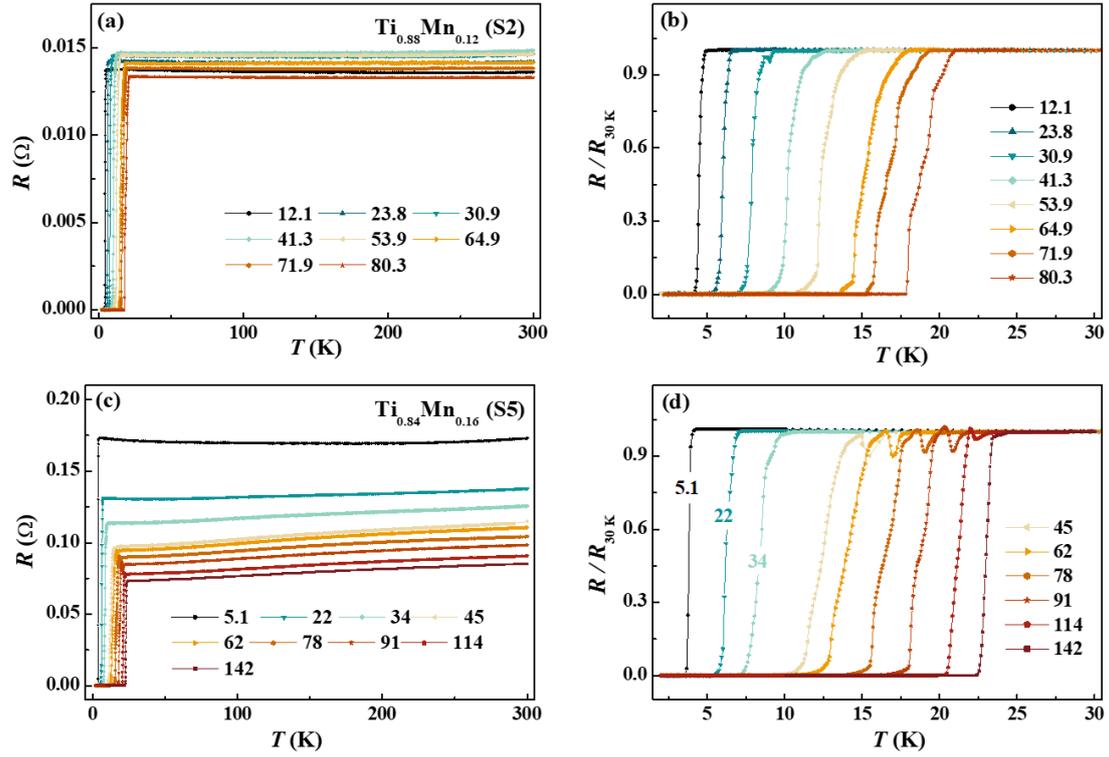

**Fig. 1 | $T_c$s of $Ti_{1-x}Mn_x$ at different pressures.** (a) Temperature-dependent electrical resistance of $Ti_{0.88}Mn_{0.12}$ from 2 to 300 K at different pressures up to 80.3 GPa. (b) Normalized resistance of $Ti_{0.88}Mn_{0.12}$ in the vicinity of superconductivity. (c) Temperature-dependent electrical resistance of $Ti_{0.84}Mn_{0.16}$ from 2 to 300 K at different pressures up to 142 GPa. (d) Normalized resistance of $Ti_{0.84}Mn_{0.16}$ in the vicinity of superconductivity.

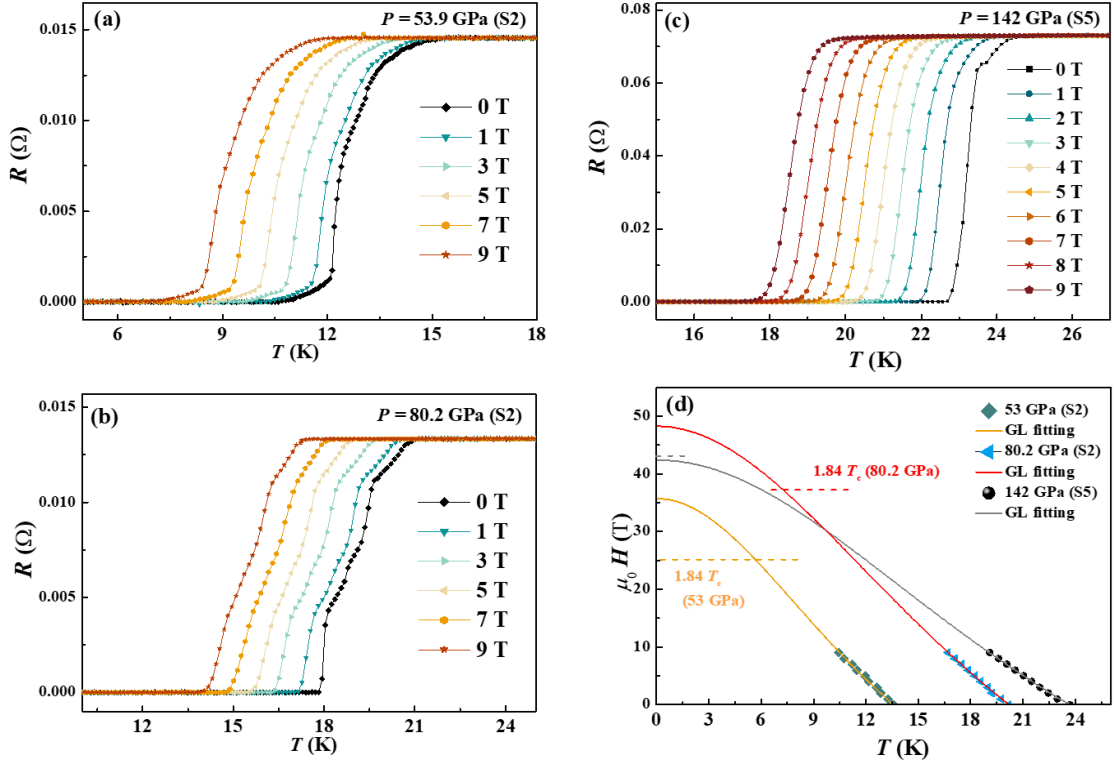

**Fig. 2 | Upper critical fields of Ti$_{1-x}$Mn$_x$ under various pressures.** (a-c) Temperature-dependent electrical resistance of Ti$_{1-x}$Mn$_x$ under different fields at (a) 53.9 GPa and (b) 80.2 GPa for S2, and (c) 142 GPa for S5. (d) Temperature dependence of the upper critical fields ($\mu_0 H_{c2}$) fitted by the Ginzburg-Landau (GL) equation; the dashed lines mark the corresponding Pauli-limiting fields ($\mu_0 H_p = 1.84 T_c$) at different pressures.

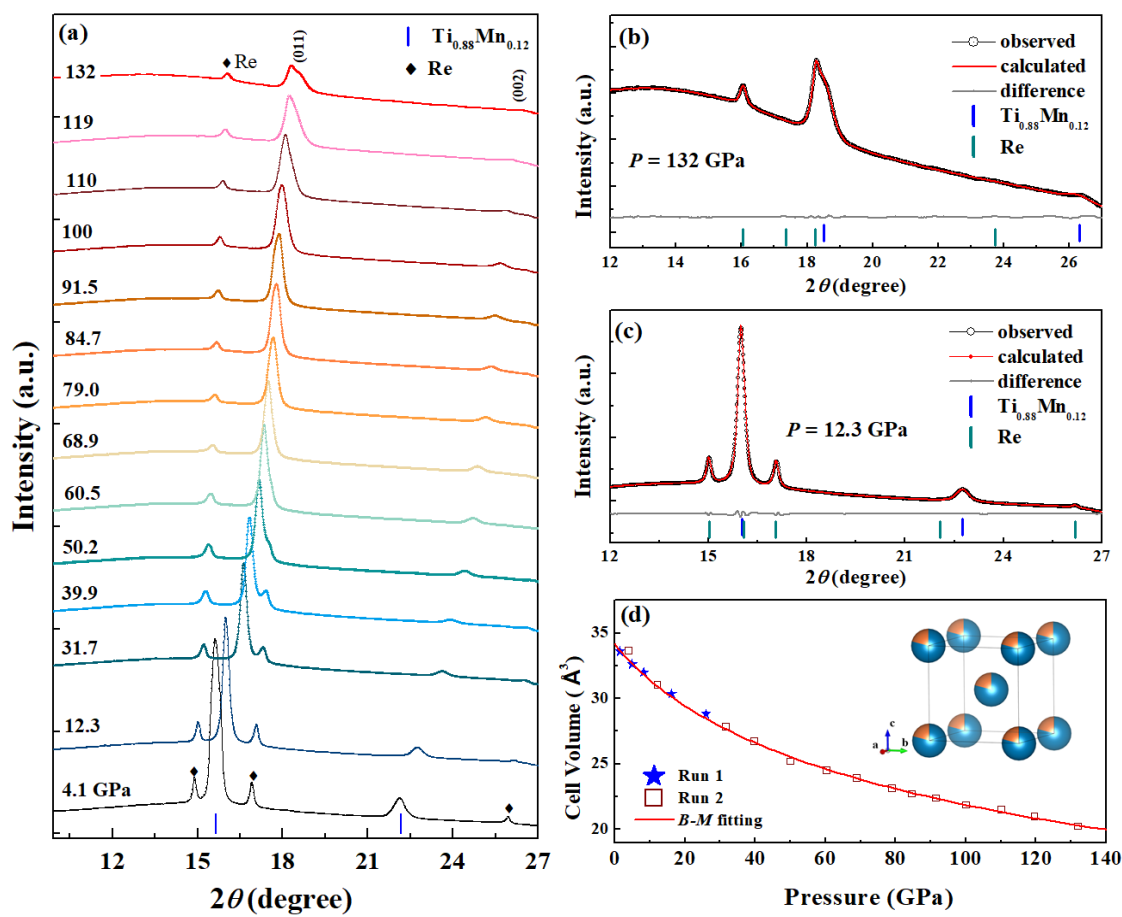

**Fig. 3 | High-pressure synchrotron X-ray diffraction on Ti$_{1-x}$Mn$_x$.** (a) The *in situ* high-pressure X-ray diffraction patterns of Ti$_{0.88}$Mn$_{0.12}$ up to 132 GPa with the wavelength of 0.6199 Å. Vertical bars and solid rhomboids represent the diffraction peaks of Ti$_{0.88}$Mn$_{0.12}$ and Re, respectively. (b,c) The typical Rietveld fitting curves with the pressure at 132 GPa and 12.3 GPa. The red solid line and black circles represent the calculated and experimental data, respectively. The vertical bars are the diffraction peak positions. (d) Pressure dependence of the cell volume. The solid red curve is the third-order Birch-Murnaghan fitting line.

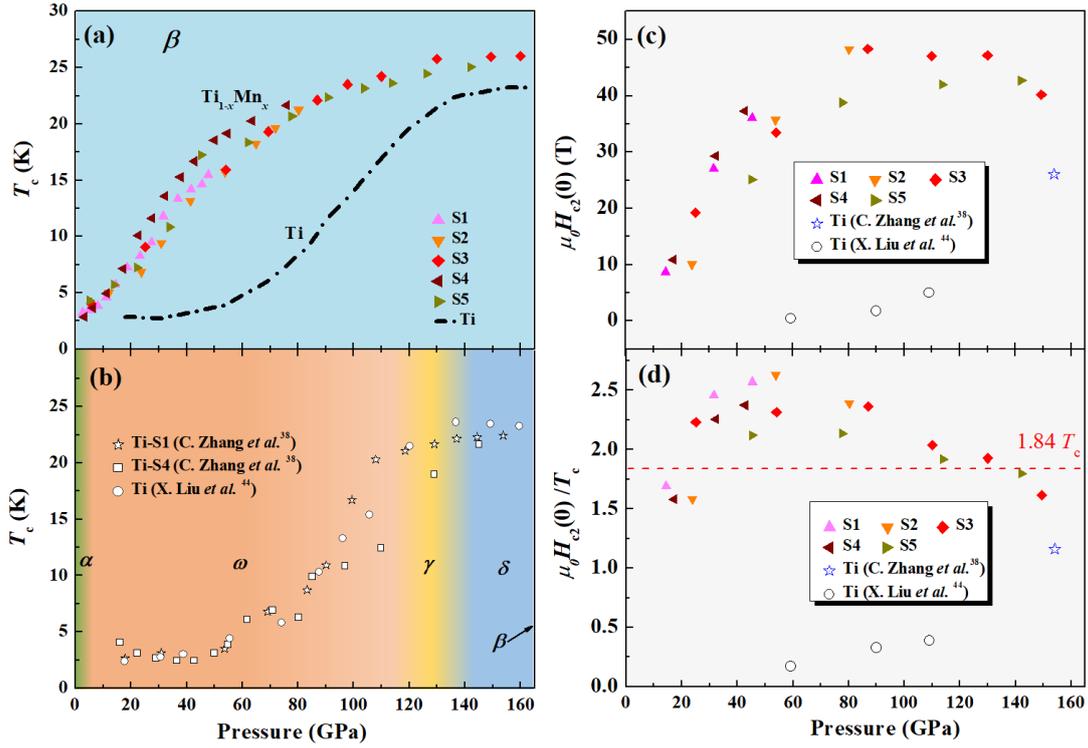

**Fig. 4 | Evolution of $T_c$ and $\mu_0 H_{c2}$ of $Ti_{1-x}Mn_x$ under high pressures.** (a) Temperature-pressure phase diagram of $Ti_{1-x}Mn_x$ with $x = 0.12$ (sample 1, 2, 3) and $x = 0.16$ (sample 4, 5), while the dashed line marks the superconducting transition temperature of Ti under high pressures. (b) Superconducting phase diagram of elemental Ti from Refs. [38, 44]. (c) Pressure-dependent $\mu_0 H_{c2}(0)$ of $Ti_{1-x}Mn_x$ for different samples. (d) Pressure dependence of $\mu_0 H_{c2}(0)/T_c$, the value of Pauli limit $\mu_0 H_p = 1.84 T_c$ is indicated by the red dashed line. The open circles and triangles in (c) and (d) represent the $\mu_0 H_{c2}(0)$ and $\mu_0 H_{c2}(0)/T_c$ of element Ti at corresponding pressures.

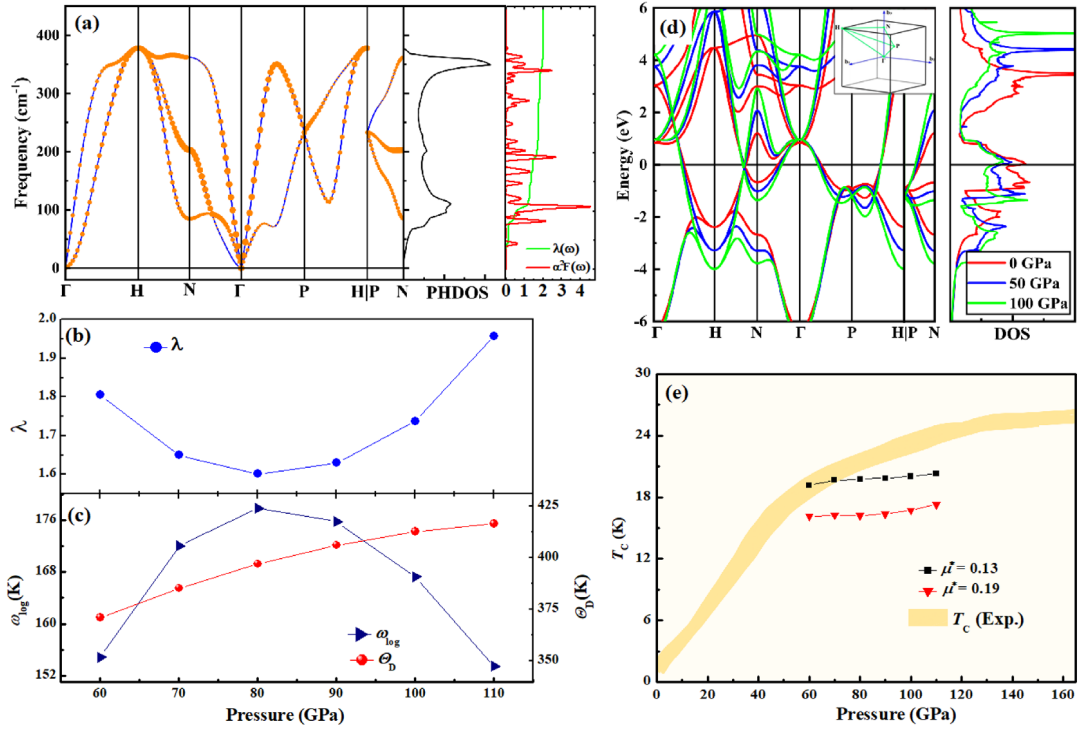

**Fig. 5 | DFT calculation on Ti$_{0.88}$Mn$_{0.12}$ under high pressures.** (a) Calculated phonon spectrum, phonon density of states, Eliashberg spectral function $\alpha^2F(\omega)$, and accumulated EPC strength $\lambda(\omega)$ at 100 GPa. The sizes of the orange dots on the phonon dispersion are proportional to the EPC strengths. (b,c) Total electron-phonon coupling parameter $\lambda$, logarithmic average frequency $\omega_{\log}$ and calculated Debye temperature $\Theta_D$ of Ti$_{0.88}$Mn$_{0.12}$ in the pressure range from 60 to 110 GPa. (d) Calculated electronic band structures and density of states of Ti$_{1-x}$Mn$_x$ at various pressures. The inset shows the Brillouin zone with green lines representing high-symmetry-point paths. (e) DFT calculation of $T_c$ versus pressure using different $\mu^*$ for Ti$_{0.88}$Mn$_{0.12}$. The yellow area represents the values of $T_c$ extracted from experimental data in Fig. 4(a).